\documentclass[%
prl,
 reprint,
 amsmath,amssymb,
 aps,preprintnumbers]{revtex4-1}

\usepackage{amssymb,amsmath}
\usepackage{textcomp}
\usepackage{tikz}
\usepackage{graphicx}
\usepackage{float}
\usepackage{textpos}
\usepackage{mathtools}
\usepackage{xcolor}
\usepackage{verbatim}

\usepackage{graphicx} 

\def\XXint#1#2#3{{\setbox0=\hbox{$#1{#2#3}{\int}$}
     \vcenter{\hbox{$#2#3$}}\kern-.5\wd0}}

\usepackage[colorlinks=true,linkcolor=black, citecolor=black,
urlcolor=black]{hyperref}
\usepackage{xcolor}

\usepackage{multirow,graphics}
    \newcommand{\beq}{\begin{equation}}
    \newcommand{\eeq}{\end{equation}}
    \newcommand\beqa{\begin{eqnarray}}
    \newcommand\eeqa{\end{eqnarray}}

\renewcommand{\leq}{\leqslant}

\renewcommand{\geq}{\geqslant}

\usepackage{amstext}
\usepackage{amssymb}
\usepackage{amsmath}
\usepackage{graphicx}
\usepackage{color}

\usepackage{textcomp}
\usepackage{tikz}
\usepackage{graphicx}
\usepackage{float}

\usepackage{xcolor}

\usepackage{esint}

\usepackage{bbm}

\begin{document}

\makeatletter
     \@ifundefined{usebibtex}{\newcommand{\ifbibtexelse}[2]{#2}} {\newcommand{\ifbibtexelse}[2]{#1}}
\makeatother

\preprint{}

\newcommand{\footnoteab}[2]{\ifbibtexelse{%
\footnotetext{#1}%
\footnotetext{#2}%
\cite{Note1,Note2}%
}{%
\newcommand{\textfootnotea}{#1}%
\newcommand{\textfootnoteab}{#2}%
\cite{thefootnotea,thefootnoteab}}}

\title{Continuous Quivers}

\author{Evgeny Sobko}
\email{evgenysobko $\bullet$ gmail.com}
\affiliation{London Institute for Mathematical Sciences, Royal Institution, London, W1S 4BS, UK}

\begin{abstract}
We consider half-BPS Wilson loops in \(\mathcal{N} = 2\) long circular quiver gauge theories at large-\(N\) with continuous limit shape of ’t Hooft couplings. In the limit of an infinite number of nodes \(L\), the solution to the localisation matrix model is given by Wigner semicircles for any profile of couplings. Higher-order corrections in \(1/L\) can be calculated iteratively. Combining large \(L\) with a strong coupling regime we identify a double scaling limit that describes dynamics along a fifth dimension which emerges dynamically from the quiver diagram. We solve the resulting integro-differential equation exactly for a certain range of parameters. 
\end{abstract}

\maketitle
\section{Introduction}
Physicists have been acquainted with quiver gauge theories since the 1980s, initially employing them as composite models \cite{Georgi:1985hf}. Later, quivers garnered increased attention, naturally arising in geometrical engineering from D-branes \cite{Douglas:1996sw}, discretised versions of space-time \cite{Arkani-Hamed:2001kyx} and as  holographic duals. 

The chrestomathic example of holographic duality is \(\mathcal{N}=4\) SYM dual to IIB string theory on \(AdS_5\times S^5\) \cite{Maldacena:1997re}. This duality has passed various nonperturbative checks, particularly matching the highly nontrivial calculations of the Wilson loop expectation value from both gauge and string theory sides. In gauge theory, the exact calculation of the BPS Wilson loop can be achieved using localization techniques \cite{Pestun:2007rz}, which reduce the problem to a finite dimensional matrix integral. In the case of \(\mathcal{N}=4\) SYM, this matrix integral is Gaussian and the Wilson loop can be calculated exactly even at finite \(N\). Another class of theories enabling exact localisation computation is four-dimensional \(\mathcal{N}=2\) circular quiver gauge theories which are dual to IIB string theory on the orbifold \(AdS_5\times (S^5/Z^L)\)\cite{Kachru:1998ys}, where \(L\) is the number of nodes in the quiver. The matrix model in this case is interacting even for equal couplings but decouples into Gaussian integrals at large-\(N\). In the case of general 't Hooft coupling constants, the saddle point approximation results in a system of \(L\) coupled integral equations for \(L\) densities of eigenvalues of scalar fields. These densities have finite supports and they are equal in the case of equal couplings. For arbitrary couplings, all supports are different, and there is no general method to solve these integral equations exactly. In the strong coupling limit, the sizes of supports go to infinity and one can use the Wiener-Hopf method to calculate the first few terms  \cite{Zarembo:2020tpf,Ouyang:2020hwd}. 

\begin{figure}
   \includegraphics[scale=0.51]{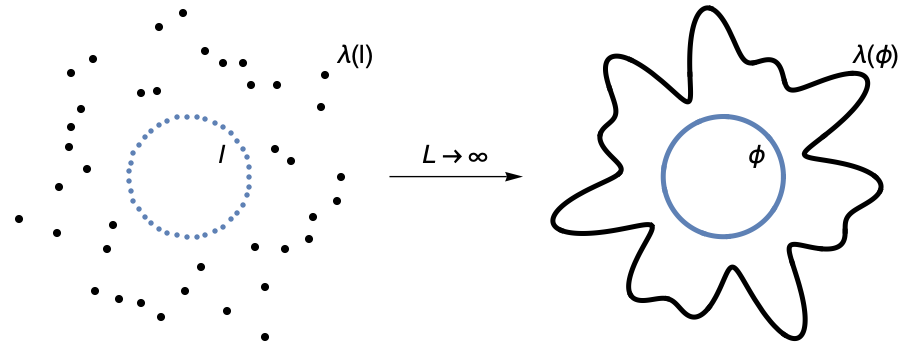}
   \label{fig:Quiver} 
\caption{The left picture schematically represents the circular quiver, with the set of 't Hooft couplings depicted by blue and black dots, respectively. The right picture illustrates their continuous large-L limit shapes.}
\end{figure}

Surprisingly, a similar system of integral equations appears in a completely different area of physics, - two-dimensional integrable quantum field theories. Integrable models can be solved using the Bethe Ansatz technique and the resulting equations in the thermodynamic limit turn into a system of coupled integral equations written on the densities of Bethe roots or quasienergies. In the presence of general external fields all supports vary, rendering the system of equations unsolvable analytically, which is analogous to the scenario of saddle point equations for quivers with general couplings discussed earlier.  In a recent paper \cite{Kazakov:2023imu} we considered  the \(SU(N)\) Principal Chiral Model (PCM) in arbitrary external fields with continuous limit shape at large-\(N\). The condition of smoothness of the limit shape turned out to be so constrained that it allowed us to find an exact solution. Being inspired by this observation, in this paper, we will consider the case of long quivers with a continuous limit shape of coupling constants, as schematically illustrated in FIG \ref{fig:Quiver}. 1, and once again it will enable us to find an exact solution. In this case the analogy with PCM goes even further: the \(A_{N-1}\) Dynkin diagram at large-\(N\) transforms into a new third dimension, while in the case of 4D quiver gauge theory, the \(A_{L-1}\) quiver diagram efficiently turnes into the fifth dimension \cite{Arkani-Hamed:2001kyx}.  
The double-scaled PCM has been conjectured to be dual to a noncritical string theory \cite{Kazakov:2019laa}, although its formulation remains unknown. Conversely, circular quiver gauge theory already possesses a known dual string description at finite
\(L\), and this provides a good starting point for understanding a dual picture of the DS limit explored in this paper.

We will calculate the expectation value of the half-BPS Wilson loop at large-\(L\) for an arbitrary continuous profile of couplings. Then, in strong coupling we will identify the DS limit sending \(L\) and the couplings to infinity in a synchronised way. For a certain range of parameters we will solve the resulting integro-differential equation exactly and conclude with a discussion of future directions.
 
\section{Circular quivers at Large \(L\)}
The field content of the \(SU(N)^{L}\) circular quiver consists of \(L\) vector multiplets in the adjoint representation, and \(L\) bi-fundamental matter hypermultiplets. In what follows we will be interested in the expectation value of the Wilson loop :
\begin{gather}
W_l=\langle \frac{1}{N}\text{Pexp}\oint\limits_{C}ds(i\dot{x}^\mu A_{l\mu}+|\dot{x}|\Phi_l) \rangle 
\end{gather}
for the circular contour \(C\).

Once the theory is compactified on the four-dimensional sphere, the expectation value of half-BPS Wilson loop can be written in the form of the matrix integral using localisation technique \cite{Pestun:2007rz}. The large-\(N\) saddle point approximation results into the system of integral equations for the densities \(\rho_l(x)\) of eigenvalues of the scalar fields \(\Phi_l\) \cite{Zarembo:2020tpf,Ouyang:2020hwd}:  
\begin{gather}
\frac{1}{2}\int\limits_{-\mu_{l-1}}^{\mu_{l-1}}dy \rho_{l-1}(y)K(x-y)+\frac{1}{2}\int\limits_{-\mu_{l+1}}^{\mu_{l+1}}dy \rho_{l+1}(y)K(x-y)\notag\\
+\int\limits_{-\mu_l}^{\mu_l}dy \rho_l(y)\left(\frac{1}{x-y}-K(x-y)\right)=\frac{8\pi^2}{\lambda_l}x
 \label{OriginalEqns}
\end{gather}
where \(\{\lambda_l\}\) is the set of ’t Hooft couplings, the kernel 
\begin{gather}
K(x)=x\left(\psi(1+i x)+\psi(1-i x)+2\gamma\right) \label{Kernel}
\end{gather}
is regular, index \(l\) numerates quiver's nodes \(l\in \{1,...,L\}\) and periodicity \(l\sim l+L\) is assumed. Densities vanish at the boundaries \(\rho_l(\pm \mu_l)=0\) and are normalised to unity:
\begin{gather}
\int\limits_{-\mu_l}^{\mu_l}dx \rho_l(x)=1. \label{Normalization}
\end{gather}
Once the densities are known, the expectaion value of the Wilson loop can be easily calculated :
\begin{gather}
W_l=\int\limits_{-\mu_l}^{\mu_l}dx \rho_l(x)e^{2\pi x}\label{ExpValDef}
\end{gather}
In what follows, there will be useful to introduce another parametrization of ’t Hooft couplings through the effective coupling constant \(\bar{\lambda}\) and the set of angle variables \(\{\theta_l\}\):
\begin{gather}
\lambda_l=\frac{2\pi\bar{\lambda}}{L \theta_l}, \ \ \bar{\lambda}^{-1}=\frac{1}{L}\sum\limits_{l=1}^L\lambda_l^{-1}, \ \ \sum\limits_{l=1}^L\theta_l=2\pi
\end{gather}

Except trivial case of equal couplings, the system of integral equations \eqref{OriginalEqns} can't be solved exactly at finite \(L\). Being inspired by \cite{Kazakov:2023imu}, in this paper we  introduce the  \textit{continuous circular quivers} such that  the number of nodes \(L\) goes to infinity and \(\lambda_l=\lambda (\frac{2\pi l}{L})\), where \(\lambda(\phi)\in C^\infty(S^1)\) is a smooth strictly positive function on the circle parametrised by the continuous coordinate \(\phi=\lim\limits_{L\rightarrow\infty}\frac{2\pi l}{L}\).

In the leading large-\(L\) limit terms with \(K\) kernel vanish:
\begin{gather}
I_{l+1}(x)+I_{l-1}(x)-2I_l(x)\stackrel{L\rightarrow \infty}{\simeq}(2\pi/L)^2\partial_\phi^2 I(\phi,x),\\
I_l(x)=\int\limits_{-\mu_l}^{\mu_l}dy \rho_l(y)K(x-y)\notag
\end{gather}
and we end up with a Gaussian matrix model :
\begin{gather}
\int\limits_{-\mu_0(\phi)}^{\mu_0(\phi)}dy \frac{\rho_0(\phi,y)}{x-y}=\frac{8\pi^2}{\lambda(\phi)}x+O(L^{-2})
\end{gather}
The solution is given by Wigner semicircles with the radii fixed by the normalisation \eqref{Normalization} :
\begin{gather}
\rho_0(\phi,x)=\frac{2}{\pi {\mu_0(\phi)}^2}\sqrt{{\mu_0(\phi)}^2-x^2}, \ \  \mu_0(\phi)=\frac{\sqrt{\lambda(\phi)}}{2\pi}
\end{gather}
We stress again that the existence of this solution crucially relies on the smoothness of the limit shape.

In order to get the next terms in \(1/L\) expansion
we use inverse formula for Cauchy kernel and rewrite the original equations in the following form :
\begin{gather}
\rho_l(x)=\sqrt{\mu_l^2-x^2}\frac{8\pi}{\lambda_l}\ +\notag\\
\sqrt{\mu_l^2-x^2}\int\limits_{-\mu_l}^{\mu_l}dt \frac{I_{l+1}(t)+I_{l-1}(t)-2I_l(t)}{2\pi^2\sqrt{\mu_l^2-t^2}(x-t)}\label{NewIntEq}
\end{gather}
The leading \(L^{-2}\) correction to the density  arises from the second term, where it suffices to substitute the leading semicircle for the exact density:
\begin{gather}
\rho(\phi,x)=\frac{2}{\pi {\mu_0(\phi)}^2}\sqrt{\mu(\phi)^2-x^2}\ +\label{LOinL_rho}\\
\frac{2}{L^2}\sqrt{\mu(\phi)^2-x^2}\int\limits_{-\mu(\phi)}^{\mu(\phi)}dt \frac{\frac{\partial^2}{\partial\phi^2}I_0(\mu(\phi),t)}{\sqrt{\mu(\phi)^2-t^2}(x-t)}+O(L^{-4}),\notag\\
I_0(\mu(\phi),x)=\int\limits_{-\mu(\phi)}^{\mu(\phi)}dy\frac{2}{\pi {\mu(\phi)}^2}\sqrt{{\mu(\phi)}^2-y^2}K(x-y).\notag
\end{gather}
The size of support is fixed by the normalization \eqref{Normalization}:
\begin{gather}
\mu(\phi)=\mu_0(\phi)\ +\label{LOinL_support}\\
\frac{\pi\mu^2_0(\phi)}{L^2}\int\limits_{-\mu_0(\phi)}^{\mu_0(\phi)}dx\frac{ \partial^2_\phi I_0(\mu_0(\phi),x)}{\sqrt{\mu_0(\phi)^2-x^2}}T_1\left(\frac{x}{\mu_0(\phi)}\right)+O(L^{-4})\notag
\end{gather}
where \(T_1(x)\) is the first Chebyshev polynomial of the first kind. 

In principle, one can go further and generate next corrections for the density and support 
\begin{gather}
\rho(\phi,x)=\sum \rho_{2k}(\phi,x)L^{-2k},\\ 
\mu(\phi)=\sum \mu_{2k}(\phi)L^{-2k}
\end{gather}
iteratively solving integro-differential equation:
\begin{gather}
\int\limits_{-\mu(\phi)}^{\mu(\phi)} dy\frac{\rho(\phi,y)}{x-y}+\frac{2\pi^2}{L^2}\partial^2_\phi\int\limits_{-\mu(\phi)}^{\mu(\phi)}dy K(x-y)\rho(\phi,y) =\frac{8\pi^2}{\lambda(\phi)}x \label{IterativeEqnLargeL}
\end{gather}
Finally one can plug \eqref{LOinL_rho} and \eqref{LOinL_support} into the \eqref{ExpValDef} and get the expectation value of the Wilson loop, however for the general profile of coupling constants the answer can be written only in the form of multiple integrals or infinite sums. At the same time, it's straightforward to expand the kernel \eqref{Kernel} at small arguments and  generate weak coupling \(\mu(\phi)\rightarrow 0\) expansion :
\begin{gather}
W(\phi)\stackrel{\bar{\lambda}\rightarrow 0}{\simeq} 1+\frac{\lambda(\phi)}{8}+O(\lambda^2)+\notag\\
+\frac{1}{L^2}\left(\frac{3\zeta(3)}{ 2^7 \pi}\lambda^2(\phi)\lambda''(\phi)+O(\lambda^4)\right)
\end{gather}
while from the expansion at large arguments we get the  strong coupling  \(\mu(\phi)\rightarrow \infty\) asymptotics : 
\begin{gather}
W(\phi)\stackrel{\bar{\lambda}\rightarrow \infty}{\simeq} \frac{e^{2\pi \mu}}{2\pi^2\mu_0^\frac{3}{2}}\left(1+\frac{\pi^2}{4L^2}(\mu_0\mu''_0-15(\mu'_0)^2) \right) \label{WLoopStrCoup}\\
\mu(\phi)\stackrel{\bar{\lambda}\rightarrow \infty}{\simeq} \mu_0\left(1+\frac{\pi^2}{2L^2}(\mu_0 \mu'_0)'+L^{-2}O(1)\right)\label{SupportStrCoup}
\end{gather}
where for sake of brevity we omitted explicit \(\phi\) dependence on the rhs.  
Also we didn't expand the exact size \(\mu(\phi)\) of support in the power, keeping in mind the next section where we will introduce the double-scaling limit.

\section{Double-scaling limit}
The leading terms in the strong coupling expansion  \eqref{WLoopStrCoup}-\eqref{SupportStrCoup} are proportional to \(\Lambda^2/L^2\) where \(\Lambda\) is a scale of support \(\mu\sim \Lambda\). Inspecting further the iterative solution of \eqref{IterativeEqnLargeL} at strong coupling  one can see that the whole expansion for the densities and supports goes over the powers of \((\Lambda/L)^2\) and subleading terms : \(\sum_k c_k(\Lambda/L)^{2k}(1+O(L^{-2},\Lambda^{-2}))\). This pattern suggests resummation of the leading subsequence in the double-scaling limit \(L,\Lambda\rightarrow\infty\) with the finite ratio \(\Lambda/L\). Namely, let's perform the following rescaling in the original equations \eqref{OriginalEqns}:
\begin{gather}
\mu(0)=\Lambda, \ \ \mu(\phi)=\Lambda \nu(\phi), \ \ x=\Lambda \alpha, \\ \lambda(\phi)=\Lambda^2 \gamma(\phi), \ \ \rho(\phi,x)=\Lambda^{-1}\varepsilon(\phi,x)
\end{gather}
sending \(L\) and \(\Lambda\) to infinity such that the ratio \(\chi^2=2\pi^2\Lambda^2/L^2\) is fixed and \(\gamma(\phi)\in C^\infty(S^1)\), we will call it \textit{DS continuous circular quiver} . It is straightfoward to take this limit and the resulting integro-differential equation has the simple form 
\begin{gather}
\int\limits_{-\nu(\phi)}^{\nu(\phi)}d\beta \frac{\varepsilon(\phi,\beta)}{\alpha-\beta}+\chi^2\partial^2_\phi\int\limits_{-\nu(\phi)}^{\nu(\phi)}K_{_{DS}}(\alpha-\beta)\varepsilon(\phi,\beta) =\frac{8\pi^2}{\gamma(\phi)}\alpha \notag\\
K_{_{DS}}(\alpha)=\alpha\log \alpha^2 \label{DSeq}
\end{gather}
with the following boundary and normalisation conditions :
\begin{gather}
\nu(0)=\nu(2\pi)=1, \ \ \nu'(0)=\nu'(2\pi), \label{PeriodicityDS}\\
\varepsilon(\phi,\pm\nu(\phi))=0, \ \ \int\limits_{-\nu(\phi)}^{\nu(\phi)}d\alpha\  \varepsilon(\phi,\alpha)=1.\label{NormalisationDS}
\end{gather}
Formally, the limit of small \(\chi\) can be again solved iteratively and the expansion matches the strong coupling expansion at large-\(L\) \eqref{SupportStrCoup} :
\begin{gather}
\nu(\phi)\stackrel{\chi^2\rightarrow 0}{\simeq}\nu_0\left(1+\frac{\chi^2}{4}(\nu_0 \nu'_0)'+O(\chi^4)\right)\label{SupportSmallChiDS}
\end{gather}
where \(\nu_0(\phi)=\sqrt{\gamma(\phi)}/2\pi\). However, in order to get the exact solution one has to perform the resummation of infinite number of iterations what potentially can lead to the asymptotic expansion. We leave this analysis for the future and instead turn to another regime of sufficiently large \(\chi\) where we can find exact solution. In the case of large \(\chi\) it is natural to expect the similarity with ordinary strong coupling at finite \(L\) and indeed beeing inspired by \cite{Zarembo:2020tpf,Ouyang:2020hwd} we write the following ansatz 
\begin{gather}
\varepsilon(\phi,\alpha)=A\left(\frac{2}{\pi}\sqrt{\nu(\phi)^2-\alpha^2}+\frac{c(\phi)\nu(\phi)^2}{\pi\sqrt{\nu(\phi)^2-\alpha^2}} \right)\label{EpsilonDS}
\end{gather}
which as we will see in a moment goes through the lhs of \eqref{DSeq}  even at finite \(\chi\). Formally, the  second term \((\nu(\phi)^2-\alpha^2)^{-1/2}\) diverges at the boundary, however one should understand it as a pointwise limit \((\nu^2-\alpha^2)^{-1/2}=\nu^{-2}\lim\limits_{k\rightarrow \infty}(\nu^2-\alpha^2)^\frac{1}{2}\frac{1-(\alpha/\nu)^{2k}}{1-(\alpha/\nu)^2}\) of functions satisfying correct boundary conditions. For the resolution of the true boundary behaviour one should go to the \(o(1)\) scale invisible in the DS approximation. 

Using normalisation condition \eqref{NormalisationDS} we get the relation between \(\nu(\phi)\) and \(c(\phi)\):
\begin{gather}
c(\phi)=\frac{1}{A\nu(\phi)^2}-1.
\end{gather}
Let's mention that the coefficient \(c(\phi)\) should be non-negative due to the positivity of the density \(\varepsilon(\phi)\geq0\), what translates into the following restriction :
\begin{gather}
\nu(\phi)\leq 1/\sqrt{A}.\label{ConditionDSexact}
\end{gather}
Evaluating integrals
\begin{gather}
\int\limits_{-\nu}^{\nu}\frac{d\beta}{\pi}\ \sqrt{\nu^2-\beta^2}K_{_{DS}}(\alpha-\beta)=
\alpha\nu^2\log\frac{\nu e^\frac{1}{2}}{2}+\frac{\alpha^3}{3}\\
\int\limits_{-\nu}^{\nu} \frac{d\beta}{\pi\sqrt{\nu^2-\beta^2}} K_{_{DS}}(\alpha-\beta)=2\alpha \log\frac{\nu e}{2}\label{IntOneOverSqrt}
\end{gather}
and taking derivative \(\partial^2_\phi\) we see that the result on the lhs of \eqref{DSeq} is proportional to \(\alpha\) and thus equation reduces to matching the prefactors on both sides :
\begin{gather}
A+\chi^2\left((\log \nu(\phi))''-A(\nu(\phi)\nu'(\phi))'\right)=\frac{4\pi^2}{\gamma(\phi)}\label{MatchingPrefactorsDS}
\end{gather}
Integrating this equation over \(\phi\) from \(0\) to \(2\pi\) and using periodicity \eqref{PeriodicityDS} we find the constant \(A\) :
\begin{gather}
A=2\pi\int\limits_0^{2\pi}\frac{d\phi}{\gamma(\phi)}=\frac{4\pi^2}{\bar{\gamma}}.
\end{gather}
where \(\bar{\gamma}=\lim_{\Lambda\rightarrow\infty}\Lambda^{-2}\bar{\lambda}\) is DS effective coupling.

The differential equation \eqref{MatchingPrefactorsDS} can be easily solved in quadratures and the integration constants are fixed by the boundary conditions  \eqref{PeriodicityDS}:
\begin{gather}
\log\nu(\phi)-\frac{2\pi^2}{\bar{\gamma}}\nu^2(\phi)=-\frac{2\pi^2}{\bar{\gamma}}+\label{DSeqn}\\
\frac{2\pi}{\bar{\gamma}\chi^2}\left(\phi\int\limits_0^{2\pi}d\phi'(\phi'-\Theta(\phi'))+2\pi\int\limits_0^{\phi}d\phi'(\Theta(\phi')-\phi')\right)\notag
\end{gather}
where we introduced continuous in \(L\rightarrow\infty\) limit monotonic function \(\Theta(\phi) : S^1\rightarrow S^1\)
\begin{gather}
\Theta(\phi=2\pi l/L)=\sum\limits_{i=1}^l\theta_i \ .
\end{gather}
The equation \eqref{DSeqn} defines the support \(\nu(\phi)\) for the given DS parameter \(\chi\) and the profile of coupling constants \(\gamma(\phi)\). Let's mention that this equation has solution only in the case when the rhs is less or equal than the maximum of the lhs. The function \(\log\nu-A\nu^2/2\) reaches its maximum at \(\nu=1/\sqrt{A}\) what leads to the same restriction \eqref{ConditionDSexact} which guarantees the positivity of the density. We expect that more general profiles \(\gamma(\phi)\) correspond to multi-cut solutions. 

\begin{figure}
   \includegraphics[scale=0.38]{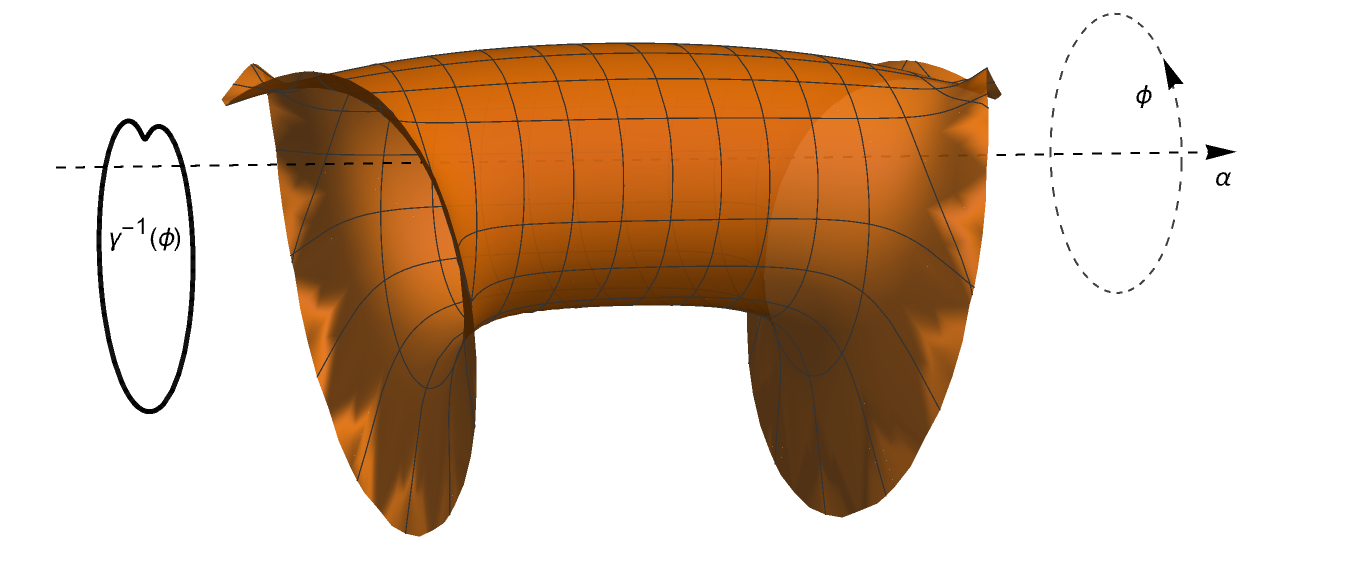}
\caption{2D surface represents the density \(\varepsilon(\phi,\alpha)\) for \(\chi^2=2\) and  \(\gamma^{-1}(\phi)=0.01(1+0.8\sin(\phi))\) drawn in the cylindric coordinates \((\varepsilon(\phi,\alpha)\cos(\phi),\varepsilon(\phi,\alpha)\sin(\phi),\alpha)\), such that \(\phi\in(0,2\pi)\), and \(\alpha\in (-0.97\nu(\phi),0.97\nu(\phi))\). We cut out the vicinity near the boundary at \(\pm 0.97\nu\) for sake of better visual representation. The black curve on the left side illustrates (scale is changed) the profile of the inverse DS coupling \(\gamma^{-1}(\phi)\).} \label{fig:DS} 
\end{figure}

The equation \eqref{DSeqn} is transcendental and in general case can be solved just numerically, see FIG. \ref{fig:DS}. In the limit of large \(\bar{\gamma}\chi^2\sim\bar{\lambda}/L^2\gg 1\) the profile of supports \(\nu(\phi)\) frozes around unity :
\begin{gather}
\nu(\phi)\stackrel{\bar{\gamma}\chi^2\rightarrow \infty}{\simeq}  1+ \label{NUtoOne}\\
\frac{2\pi}{(1-A)\bar{\gamma}\chi^2}\left(\phi\int\limits_0^{2\pi}d\phi'(\phi'-\Theta(\phi'))+2\pi\int\limits_0^{\phi}d\phi'(\Theta(\phi')-\phi')\right)\notag
\end{gather}
and the same asymptotics is valid for the limit of equal DS couplings \(\Theta(\phi)\rightarrow\phi\) at finite \(\bar{\gamma}\chi^2\). In order to reproduce the \(\nu(\phi)=1\) limit one has to send \(\gamma(\phi)\rightarrow 4\pi^2\) such that \(\Theta(\phi)\) goes to \(\phi\) faster than \(A\) goes to \(1\) so the second term in \eqref{NUtoOne} is small. In this regime \(c(\phi)\rightarrow0\) and \(\varepsilon(\phi,\alpha)\) turnes into the Wigner semicircle. However let's mention that in DS regime even for small but finite difference \(\gamma(\phi)-4\pi^2\) both terms in \eqref{EpsilonDS} are present. In order to reproduce the limit of equal original supports \(\mu(\phi)=const\) one has to leave DS regime where the difference of supports is \(\mu(\phi_1)-\mu(\phi_2)=O(\Lambda)\) and go through the completely different regime of \(\mu(\phi_1)-\mu(\phi_2)=O(1)\).  The last one is potentially accessible to Wiener-Hopf method and the corresponding solution  will be continuosly  connected  to the Wigner semicircles at equal couplings \cite{Zarembo:2020tpf,Ouyang:2020hwd}. Let's stress that the Wiener-Hopf method can't be matched directly with DS limit  due to its inapplicability to the configurations with supports \(\mu(\phi)=O(\Lambda)\). It would be very interesting to find more general solution interpolating between DS and WH regimes. 

The leading assymptotics coming from the Wiener-Hopf method \cite{Zarembo:2020tpf,Ouyang:2020hwd} functionally looks similar to \eqref{EpsilonDS}
however the terms have different scales, indeed in the ordinary strong coupling at finite \(L\) the second term is \(1/\Lambda\) suppresed in the bulk wrt the first one, and they both are of the same order  in the \(O(1)\) vicinity of the boundary. Conversely, in the DS regime both terms in the density \eqref{EpsilonDS} are of the same order in the bulk and the first one is \(1/\Lambda\) suppressed near the boundary. The last means that the leading contribution to the expectation value of the Wilson loop in DS limit comes from the second term and we get :
\begin{gather}
W^{DS}(\phi)\approx \frac{e^{2\pi \Lambda \nu(\phi)}}{\sqrt{\Lambda}}\ \frac{\frac{1}{A}-\nu^2(\phi)}{2\pi\sqrt{\nu(\phi)}}
\end{gather}

\section{Discussion}
In this paper, we considered continuous circular quivers and demonstrated that the smoothness along the quiver diagram allows for the calculation of the expectaion value of the Wilson loop at finite couplings. In the strong coupling regime we identified the double scaling limit \eqref{DSeq} and solved the corresponding equation exactly  for a certain range of parameters \eqref{ConditionDSexact}. We expect that the more general profiles correspond to the multi-cut solutions. The regimes with  \(\mu/L\ll 1\) and \(\mu/L\gg 1\) demonstrate qualitatively different behaviour. In the first case the solution has a local dependence on the quiver variable \(\phi\) \eqref{WLoopStrCoup}, \eqref{SupportStrCoup} while in the second case the density \eqref{EpsilonDS} is expressed through the functions integrated along the whole quiver \eqref{DSeqn}. 

It would be interesting to understand this picture at the microscopic level of QFT, by taking the large \(L\) limit directly at the level of the Lagrangian/lattice and analysing the emerging 5D theory \cite{Arkani-Hamed:2001kyx}.  

Effectively one can think about smoothness as an extra  constraint leading to a drastic simplification which we observed in this paper.
It would be natural to go beyond Wilson loops and study continuous limits for other quantities accessible through localisation such as  integrated correlation functions of local operators , correlators between local operators and a Wilson loop, Wilson loops in higher representations etc. Also, it would be interesting to relax the strict limit of large-\(N\) and calculate \(1/N\) corrections \cite{Beccaria:2023qnu}. Even  further, one could examine the original matrix model to check for the existence of other nontrivial regimes by synchronizing the rank of the gauge groups, representations, coupling constants, number of quiver nodes, etc.

Finally, it would be very intriguing to understand the dual string description. The strongly coupled circular quiver at finite \(L\) is  dual to a IIB string theory on the orbifold \(AdS_5\times (S^5/Z^L)\) \cite{Kachru:1998ys}.  The large \(L\) limit would take \(Z^L\) to \(U(1)\) which naively reduces dimensionality of the string theory by one to a noncritical value. On the other hand, in the continuous limit, another circle parametrized by \(\Theta(\phi)\) emerges from the quiver diagram. The parameters \(\theta_l\) are proportional to the B-fluxes through the collapsing cycles of the orbifold \cite{Lawrence:1998ja,Klebanov:1999rd} and in the DS limit they condense into a new \(S^1\) parametrized by \(\Theta(\phi)\) which can signal a nontrivial transformation of geometry. In the T-dual picture \cite{Rey:2010ry} angle variables \(\theta_i\) are  coordinates of \(L\) NS5 branes placed along \(S^1\) and the function \(\Theta(\phi)\) describes their continuous distribution in the DS limit.


We will return to these questions in future  publications.

\begin{acknowledgments}
\section*{Acknowledgments}
\label{sec:acknowledgments}

Author thanks Nadav Drukker and Gregory Korchemsky for the interesting discussions and especially Konstantin Zarembo for the illuminating  discussions at various stages of the propject, comments on the manuscript, and joint related work.
\end{acknowledgments}

\bibliography{ContinuousQuivers.bib}

\begin{thebibliography}{14}%
\makeatletter
\providecommand \@ifxundefined [1]{%
 \@ifx{#1\undefined}
}%
\providecommand \@ifnum [1]{%
 \ifnum #1\expandafter \@firstoftwo
 \else \expandafter \@secondoftwo
 \fi
}%
\providecommand \@ifx [1]{%
 \ifx #1\expandafter \@firstoftwo
 \else \expandafter \@secondoftwo
 \fi
}%
\providecommand \natexlab [1]{#1}%
\providecommand \enquote  [1]{``#1''}%
\providecommand \bibnamefont  [1]{#1}%
\providecommand \bibfnamefont [1]{#1}%
\providecommand \citenamefont [1]{#1}%
\providecommand \href@noop [0]{\@secondoftwo}%
\providecommand \href [0]{\begingroup \@sanitize@url \@href}%
\providecommand \@href[1]{\@@startlink{#1}\@@href}%
\providecommand \@@href[1]{\endgroup#1\@@endlink}%
\providecommand \@sanitize@url [0]{\catcode `\\12\catcode `\$12\catcode `\&12\catcode `\#12\catcode `\^12\catcode `\_12\catcode `\%12\relax}%
\providecommand \@@startlink[1]{}%
\providecommand \@@endlink[0]{}%
\providecommand \url  [0]{\begingroup\@sanitize@url \@url }%
\providecommand \@url [1]{\endgroup\@href {#1}{\urlprefix }}%
\providecommand \urlprefix  [0]{URL }%
\providecommand \Eprint [0]{\href }%
\providecommand \doibase [0]{http://dx.doi.org/}%
\providecommand \selectlanguage [0]{\@gobble}%
\providecommand \bibinfo  [0]{\@secondoftwo}%
\providecommand \bibfield  [0]{\@secondoftwo}%
\providecommand \translation [1]{[#1]}%
\providecommand \BibitemOpen [0]{}%
\providecommand \bibitemStop [0]{}%
\providecommand \bibitemNoStop [0]{.\EOS\space}%
\providecommand \EOS [0]{\spacefactor3000\relax}%
\providecommand \BibitemShut  [1]{\csname bibitem#1\endcsname}%
\let\auto@bib@innerbib\@empty
\bibitem [{\citenamefont {Georgi}(1986)}]{Georgi:1985hf}%
  \BibitemOpen
  \bibfield  {author} {\bibinfo {author} {\bibfnamefont {H.}~\bibnamefont {Georgi}},\ }\href {\doibase 10.1016/0550-3213(86)90092-1} {\bibfield  {journal} {\bibinfo  {journal} {Nucl. Phys. B}\ }\textbf {\bibinfo {volume} {266}},\ \bibinfo {pages} {274} (\bibinfo {year} {1986})}\BibitemShut {NoStop}%
\bibitem [{\citenamefont {Douglas}\ and\ \citenamefont {Moore}(1996)}]{Douglas:1996sw}%
  \BibitemOpen
  \bibfield  {author} {\bibinfo {author} {\bibfnamefont {M.~R.}\ \bibnamefont {Douglas}}\ and\ \bibinfo {author} {\bibfnamefont {G.~W.}\ \bibnamefont {Moore}},\ }\href@noop {} {\  (\bibinfo {year} {1996})},\ \Eprint {http://arxiv.org/abs/hep-th/9603167} {arXiv:hep-th/9603167} \BibitemShut {NoStop}%
\bibitem [{\citenamefont {Arkani-Hamed}\ \emph {et~al.}(2001)\citenamefont {Arkani-Hamed}, \citenamefont {Cohen},\ and\ \citenamefont {Georgi}}]{Arkani-Hamed:2001kyx}%
  \BibitemOpen
  \bibfield  {author} {\bibinfo {author} {\bibfnamefont {N.}~\bibnamefont {Arkani-Hamed}}, \bibinfo {author} {\bibfnamefont {A.~G.}\ \bibnamefont {Cohen}}, \ and\ \bibinfo {author} {\bibfnamefont {H.}~\bibnamefont {Georgi}},\ }\href {\doibase 10.1103/PhysRevLett.86.4757} {\bibfield  {journal} {\bibinfo  {journal} {Phys. Rev. Lett.}\ }\textbf {\bibinfo {volume} {86}},\ \bibinfo {pages} {4757} (\bibinfo {year} {2001})},\ \Eprint {http://arxiv.org/abs/hep-th/0104005} {arXiv:hep-th/0104005} \BibitemShut {NoStop}%
\bibitem [{\citenamefont {Maldacena}(1998)}]{Maldacena:1997re}%
  \BibitemOpen
  \bibfield  {author} {\bibinfo {author} {\bibfnamefont {J.~M.}\ \bibnamefont {Maldacena}},\ }\href {\doibase 10.4310/ATMP.1998.v2.n2.a1} {\bibfield  {journal} {\bibinfo  {journal} {Adv. Theor. Math. Phys.}\ }\textbf {\bibinfo {volume} {2}},\ \bibinfo {pages} {231} (\bibinfo {year} {1998})},\ \Eprint {http://arxiv.org/abs/hep-th/9711200} {arXiv:hep-th/9711200} \BibitemShut {NoStop}%
\bibitem [{\citenamefont {Pestun}(2012)}]{Pestun:2007rz}%
  \BibitemOpen
  \bibfield  {author} {\bibinfo {author} {\bibfnamefont {V.}~\bibnamefont {Pestun}},\ }\href {\doibase 10.1007/s00220-012-1485-0} {\bibfield  {journal} {\bibinfo  {journal} {Commun. Math. Phys.}\ }\textbf {\bibinfo {volume} {313}},\ \bibinfo {pages} {71} (\bibinfo {year} {2012})},\ \Eprint {http://arxiv.org/abs/0712.2824} {arXiv:0712.2824 [hep-th]} \BibitemShut {NoStop}%
\bibitem [{\citenamefont {Kachru}\ and\ \citenamefont {Silverstein}(1998)}]{Kachru:1998ys}%
  \BibitemOpen
  \bibfield  {author} {\bibinfo {author} {\bibfnamefont {S.}~\bibnamefont {Kachru}}\ and\ \bibinfo {author} {\bibfnamefont {E.}~\bibnamefont {Silverstein}},\ }\href {\doibase 10.1103/PhysRevLett.80.4855} {\bibfield  {journal} {\bibinfo  {journal} {Phys. Rev. Lett.}\ }\textbf {\bibinfo {volume} {80}},\ \bibinfo {pages} {4855} (\bibinfo {year} {1998})},\ \Eprint {http://arxiv.org/abs/hep-th/9802183} {arXiv:hep-th/9802183} \BibitemShut {NoStop}%
\bibitem [{\citenamefont {Zarembo}(2020)}]{Zarembo:2020tpf}%
  \BibitemOpen
  \bibfield  {author} {\bibinfo {author} {\bibfnamefont {K.}~\bibnamefont {Zarembo}},\ }\href {\doibase 10.1007/JHEP06(2020)055} {\bibfield  {journal} {\bibinfo  {journal} {JHEP}\ }\textbf {\bibinfo {volume} {06}},\ \bibinfo {pages} {055} (\bibinfo {year} {2020})},\ \Eprint {http://arxiv.org/abs/2003.00993} {arXiv:2003.00993 [hep-th]} \BibitemShut {NoStop}%
\bibitem [{\citenamefont {Ouyang}(2021)}]{Ouyang:2020hwd}%
  \BibitemOpen
  \bibfield  {author} {\bibinfo {author} {\bibfnamefont {H.}~\bibnamefont {Ouyang}},\ }\href {\doibase 10.1007/JHEP02(2021)178} {\bibfield  {journal} {\bibinfo  {journal} {JHEP}\ }\textbf {\bibinfo {volume} {02}},\ \bibinfo {pages} {178} (\bibinfo {year} {2021})},\ \Eprint {http://arxiv.org/abs/2011.03531} {arXiv:2011.03531 [hep-th]} \BibitemShut {NoStop}%
\bibitem [{\citenamefont {Kazakov}\ \emph {et~al.}(2024)\citenamefont {Kazakov}, \citenamefont {Sobko},\ and\ \citenamefont {Zarembo}}]{Kazakov:2023imu}%
  \BibitemOpen
  \bibfield  {author} {\bibinfo {author} {\bibfnamefont {V.}~\bibnamefont {Kazakov}}, \bibinfo {author} {\bibfnamefont {E.}~\bibnamefont {Sobko}}, \ and\ \bibinfo {author} {\bibfnamefont {K.}~\bibnamefont {Zarembo}},\ }\href {\doibase 10.1103/PhysRevLett.132.141602} {\bibfield  {journal} {\bibinfo  {journal} {Phys. Rev. Lett.}\ }\textbf {\bibinfo {volume} {132}},\ \bibinfo {pages} {141602} (\bibinfo {year} {2024})},\ \Eprint {http://arxiv.org/abs/2312.04801} {arXiv:2312.04801 [hep-th]} \BibitemShut {NoStop}%
\bibitem [{\citenamefont {Kazakov}\ \emph {et~al.}(2020)\citenamefont {Kazakov}, \citenamefont {Sobko},\ and\ \citenamefont {Zarembo}}]{Kazakov:2019laa}%
  \BibitemOpen
  \bibfield  {author} {\bibinfo {author} {\bibfnamefont {V.}~\bibnamefont {Kazakov}}, \bibinfo {author} {\bibfnamefont {E.}~\bibnamefont {Sobko}}, \ and\ \bibinfo {author} {\bibfnamefont {K.}~\bibnamefont {Zarembo}},\ }\href {\doibase 10.1103/PhysRevLett.124.191602} {\bibfield  {journal} {\bibinfo  {journal} {Phys. Rev. Lett.}\ }\textbf {\bibinfo {volume} {124}},\ \bibinfo {pages} {191602} (\bibinfo {year} {2020})},\ \Eprint {http://arxiv.org/abs/1911.12860} {arXiv:1911.12860 [hep-th]} \BibitemShut {NoStop}%
\bibitem [{\citenamefont {Beccaria}\ and\ \citenamefont {Korchemsky}(2024)}]{Beccaria:2023qnu}%
  \BibitemOpen
  \bibfield  {author} {\bibinfo {author} {\bibfnamefont {M.}~\bibnamefont {Beccaria}}\ and\ \bibinfo {author} {\bibfnamefont {G.~P.}\ \bibnamefont {Korchemsky}},\ }\href {\doibase 10.1007/JHEP04(2024)054} {\bibfield  {journal} {\bibinfo  {journal} {JHEP}\ }\textbf {\bibinfo {volume} {04}},\ \bibinfo {pages} {054} (\bibinfo {year} {2024})},\ \Eprint {http://arxiv.org/abs/2312.03836} {arXiv:2312.03836 [hep-th]} \BibitemShut {NoStop}%
\bibitem [{\citenamefont {Lawrence}\ \emph {et~al.}(1998)\citenamefont {Lawrence}, \citenamefont {Nekrasov},\ and\ \citenamefont {Vafa}}]{Lawrence:1998ja}%
  \BibitemOpen
  \bibfield  {author} {\bibinfo {author} {\bibfnamefont {A.~E.}\ \bibnamefont {Lawrence}}, \bibinfo {author} {\bibfnamefont {N.}~\bibnamefont {Nekrasov}}, \ and\ \bibinfo {author} {\bibfnamefont {C.}~\bibnamefont {Vafa}},\ }\href {\doibase 10.1016/S0550-3213(98)00495-7} {\bibfield  {journal} {\bibinfo  {journal} {Nucl. Phys. B}\ }\textbf {\bibinfo {volume} {533}},\ \bibinfo {pages} {199} (\bibinfo {year} {1998})},\ \Eprint {http://arxiv.org/abs/hep-th/9803015} {arXiv:hep-th/9803015} \BibitemShut {NoStop}%
\bibitem [{\citenamefont {Klebanov}\ and\ \citenamefont {Nekrasov}(2000)}]{Klebanov:1999rd}%
  \BibitemOpen
  \bibfield  {author} {\bibinfo {author} {\bibfnamefont {I.~R.}\ \bibnamefont {Klebanov}}\ and\ \bibinfo {author} {\bibfnamefont {N.~A.}\ \bibnamefont {Nekrasov}},\ }\href {\doibase 10.1016/S0550-3213(00)00016-X} {\bibfield  {journal} {\bibinfo  {journal} {Nucl. Phys. B}\ }\textbf {\bibinfo {volume} {574}},\ \bibinfo {pages} {263} (\bibinfo {year} {2000})},\ \Eprint {http://arxiv.org/abs/hep-th/9911096} {arXiv:hep-th/9911096} \BibitemShut {NoStop}%
\bibitem [{\citenamefont {Rey}\ and\ \citenamefont {Suyama}(2011)}]{Rey:2010ry}%
  \BibitemOpen
  \bibfield  {author} {\bibinfo {author} {\bibfnamefont {S.-J.}\ \bibnamefont {Rey}}\ and\ \bibinfo {author} {\bibfnamefont {T.}~\bibnamefont {Suyama}},\ }\href {\doibase 10.1007/JHEP01(2011)136} {\bibfield  {journal} {\bibinfo  {journal} {JHEP}\ }\textbf {\bibinfo {volume} {01}},\ \bibinfo {pages} {136} (\bibinfo {year} {2011})},\ \Eprint {http://arxiv.org/abs/1001.0016} {arXiv:1001.0016 [hep-th]} \BibitemShut {NoStop}%
\end{thebibliography}%

\end{document}